\documentstyle[epsf,fleqn,aps]{revtex}

\input epsf

\def\Section#1{}
\def\beq{\begin{equation}}
\def\eeq{\end{equation}}
\def\bea{\begin{eqnarray}}
\def\eea{\end{eqnarray}}

\begin{document}

\tolerance50000
\twocolumn[\hsize\textwidth\columnwidth\hsize\csname
@twocolumnfalse\endcsname

\title{Interplay between quasi-periodicity and disorder
in quantum spin chains in a magnetic field}

\author{M.\ Arlego}
\address{Departamento de F\'{\i}sica, Universidad Nacional de La Plata,
C.C.   67, (1900) La Plata, Argentina.
}

\maketitle

\begin{abstract}
We study the interplay between disorder and a quasi periodic
coupling array in an external magnetic field in a
spin-$\frac{1}{2}$ $XXZ$ chain. A simple real space decimation
argument is used to estimate the magnetization values where
plateaux show up. The latter are in good agreement with exact
diagonalization results on fairly long $XX$ chains.
Spontaneous susceptibility
properties are also studied, finding a logarithmic behaviour
similar to the homogeneously disordered case.
\vskip 0.5cm

\end{abstract}

\vskip -0.2cm \vskip2pc]

\vspace{.3cm}

Since their discovery in 1984 \cite{SBGC}, the properties of
quasi-crystals have been a source of sustained interest. Many
theoretical efforts on Ising models in Penrose lattices \cite{GLO}
and $XY$ Fibonacci spin chains \cite {ALM,LN,KST} have revealed
interesting magnetic orderings associated to the quasi-periodicity
of these structures. Such kind of spin arrays have been found in
recently synthesized rare earth (${\cal R}$) ZnMg-${\cal R}$
quasi-crystals (see {\it e.g.} \cite{Setal}) whose ${\cal R}$
elements have well localized $4f$ magnetic moments. The study of
quasiperiodic 1D chains has recently received renewed attention
\cite{VMG,H} and interesting properties have been elucidated. In
\cite{VMG} a system of spinless fermions in a quasiperiodic
lattice potential was studied within perturbation theory, where it
was shown that its behaviour is different from both the periodic
and the disordered cases: While in the case of a periodic
potential one may have a metal-insulator transition only if the
potential is commensurate, in the disordered case, the potential
is relevant irrespective of the position of the Fermi level. In
the quasiperiodic case, two different situations arise, depending
on whether the Fermi level coincides with one of the main
frequencies of the Fourier spectrum of the quasiperiodic potential
or not. In the first case, the situation turns out to be similar
to the periodic case while in the second, at a perturbative level,
the metal-insulator transition point is strongly modified. These
predictions have been also verified numerically in \cite{H}.
Motivated by these studies we have recently analyzed the effect of
an external magnetic field in a quasiperiodic spin chain, and
found that the magnetization curve has a very interesting nature
that could be predicted using a decimation procedure and Abelian
bosonization \cite{nos1}. For the Fibonacci case, in particular,
one can reproduce the main plateaux within the bosonization
approach by approximating the quasiperiodic modulation by
considering a subset of the main Fourier frequencies. From the
experimental point of view, interest of quasiperiodic systems
arise from artificially grown quasi-periodic heterostructures
\cite{Merlin}, quantum dot crystals \cite {Kouwen} and magnetic
multilayers \cite{BACA}.

In this short note we go one step further to analyze the interplay
between a quasiperiodic array of couplings and disorder in a $XXZ$
spin chain in the presence of an external magnetic field. Using a
simple decimation procedure we predict the appearance of plateaux
in the magnetization curve at values of $M$ which depend on both
the quasiperiodicity and the strength of the disorder. As in the
case of $p$-merized chains the presence of binary disorder results
in a shift on the plateaux positions as a function its strength,
while for the Gaussian case the plateaux are wiped out. We have
also studied the effects of a binary disorder on the double
frequency $XX$ chain studied in \cite{nos1}, where it turns out
that the plateaux structure of the pure case disappears. We have
checked this behaviour by studying numerically fairly long $XX$
systems. We also predict a logarithmic behaviour of the susceptibility at
low fields by extending the arguments in \cite{FDM} and have also
verified this behaviour numerically. Other authors have also
studied random spin systems recently \cite{PR}-\cite{Tnew}.

Let us consider the antiferromagnetic system

\begin{eqnarray}
H_{qp}=J &&\sum_{n}\left( 1+\epsilon _{n}\right) \,\left(
\,S_{n}^{x}\,S_{n+1}^{x}\,+\,S_{n}^{y}\,S_{n+1}^{y}\right.
\nonumber \\ &+&\left. \Delta \,S_{n}^{z}\,S_{n+1}^{z}\,\right)
\,-\,h\,\sum_{n}S_{n}^{z}\,,  \label{qp}
\end{eqnarray}
where $S^{x},S^{y},S^{z}\,$ denote the standard spin-$\frac{1}{2}$
matrices, in a magnetic field $h$ applied along the anisotropy
direction $(\,|\Delta |\leq 1\,)$. Here, the coupling modulation
is introduced via the $\epsilon _{n}$ parameters defined as
$\epsilon _{n}=\sum_{\nu }\delta _{\nu }\,\cos \left( 2\pi
\,\omega _{\nu }n\,\right) \,$, so quasi-periodicity arises upon
choosing an irrational subset of frequencies $\omega _{\nu }$ with
amplitudes $\delta _{\nu }\,$.

Furthermore, the couplings $J_n$ are randomly distributed.
Specifically, we consider a binary distribution of strength $p$
($p=0$ corresponds to the pure quasiperiodic case while $p=1$
corresponds to the uniform chain),

\begin{equation}
P(J_{i})=p\ \delta (J_{i}-U)+(1-p)\ \delta (J_{i}-J(1+\epsilon _{n}))\
,
\label{q-bin}
\end{equation}
with $\epsilon _{n}$ defined as above, along with a Gaussian
disorder $ P(J_{i})\propto \exp
-\frac{(J_{i}-\overline{J_{i}})^{2}}{2\,\sigma _{i}^{2}} $. These
distributions, taken with same mean and variance, are built to
enforce quasiperiodicity. Thus, on average $\epsilon _{n}$ is a
measure of the couplings quasiperiodicity. In what follows we
assume that $U$ is the smallest coupling

We will follow the decimation procedure as described in \cite{FDM}
to obtain the value of the magnetization for the main plateaux. In
our problem (which is at $T=0$) the energy scale is provided by
the magnetic field, and in order to compute the magnetization,
decimation has to be stopped at an energy scale of the order of
the magnetic field. We assume that all spins coupled by bonds
stronger than the magnetic field form singlets and do not
contribute to the magnetization, whereas spins coupled by weaker
bonds are completely polarized. The magnetization is thus
proportional to the fraction of remaining spins at the step where
we stop decimation. This simple argument happens to apply well to
the binary distribution, provided the energy scales of the
involved exchanges are well separated.

In studying irrational frequencies or other quasiperiodic
modulations, it is natural to analyze the case of the Fibonacci
chain, a coupling array $ J_{A}=J\left( 1+\delta \right) $,
$J_{B}=J\left( 1-\delta \right) $ generated by iterating the
substitution rules $B\rightarrow A$ and $ A\rightarrow AB$
\cite{KST}, \cite{VMG}, \cite{nos1}, \cite{AchiamLubMarsh}; with the
distribution (\ref{q-bin}) $P(J_{i})=p\ \delta (J_{i}-U)+(1-p)\
\delta (J_{i}-J_{A,B})$.

We evaluate by decimation the magnetization of the widest plateaux
in the {\it strong} coupling limit $\left( \delta \rightarrow \pm
1\right)$. There are two different cases to consider, according to
$\delta \approx -1$, i.e., $J_{B}\gg J_{A}$, and the opposite
situation for $\delta \approx 1$.

Starting from saturation, in the first case the magnetic field is
lowered until it reaches the value $h_{c}\approx J_{B}$ at which
the type-{\it B} bonds experience a transition from the state of
maximum polarization to the singlet state. The magnetization at
this plateau is then obtained by decimating the {\it B} bonds.
This yields:

\begin{equation}
\left\langle M\right\rangle =1-2\frac{N_{B}}{N_{T}}=1-2\left( 1-p\right)
\frac{1}{\gamma ^{2}} \ ,  \label{decB>A}
\end{equation}
where $N_{T}=N_U+N_{A}+N_{B}$ denotes the total number of bonds, $
N_{A,B}$ the number of {\it A} and {\it B} bonds respectively and
$\gamma^2 = (N_A + N_B)/N_B $. For a large iteration number of the
rules referred to above $\left( N_T\rightarrow \infty \right) $,
$N_{A}/N_{B}$ approaches the golden mean $\gamma =\left( \frac{1+
\sqrt{5}}{2}\right)$. In the $p=0$ limit, we recover the results
in \cite{nos1} and a non-vanishing $p$ results in a shift of the
position of the plateau.

In the second case, $J_{A}\gg J_{B}$, we have to distinguish two
different unit cells since type-{\it A} bonds can appear either in
pairs (forming trimers) or isolated (forming dimers). It can be
readily checked that when lowering the magnetic field from
saturation the first spins to be decimated correspond to those
forming trimers. We then have a plateau (the nearest to
saturation) at:

\begin{equation}
\left\langle M\right\rangle_{1} =1-2\frac{N_{AA}}{N_{T}}=1-2\left(
1-p\right) ^{2}\frac{1}{\gamma ^{3}},  \label{dec1A>B}
\end{equation}
where $N_{AA}$ refers to {\it A} pairs. The second plateau is
obtained after decimation the type-{\it A} bonds, and then we must
consider all the sequences $J_{A}$ between the others bonds. That
gives for the second plateau:

\begin{eqnarray}
\left\langle M\right\rangle_{2} &=&\left\langle M_{1}\right\rangle -2\left[
\left( 1-p\right) ^{3}\frac{1}{\gamma ^{4}}+\right.  \nonumber \\
&&\left. 2p\left( 1-p\right) ^{2}\frac{1}{\gamma ^{2}}+p^{2}\left(
1-p\right) \frac{1}{\gamma }\right]\ .  \label{dec2A>B}
\end{eqnarray}
Again, we recover our results in \cite{nos1} for $p=0$.

With this simple technique, one can predict the presence and
position of the plateaux, provided that there is a finite
difference between the highest values of the couplings in the
inequivalent sites.

Since the decimation procedure applies for generic XXZ chains
\cite{FDM}, \cite{nos2}, we conclude that the emergence of these strong
coupling plateaux is a generic feature, at least with an
antiferromagnetic anisotropy parameter $ 0 < \Delta < 1 $.

\begin{figure}[tbp]
\hbox{%
\epsfxsize=3.5in
\epsffile{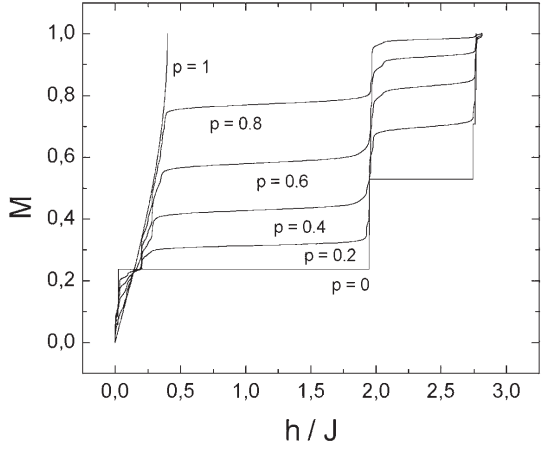}}
\caption{Magnetization curves of modulated XX Fibonacci spin
chains, immersed in disordered binary backgrounds of strength $p$
after averaging over $5\times 10^{4}$ samples with $f(18)=2584$
sites, $\protect\delta =0.95$ , $U=0.2$ and
$p=0,0.2,0.4,0.6,0.8,1$ in ascending order. The left and rightmost
lines denote respectively the pure uniform and pure Fibonacci
cases.} \label{fig1}
\end{figure}

To enable an independent check of these assertions, we turn to a
numerical diagonalization of the Hamiltonian (\ref{qp}) contenting
ourselves with the analysis of the particular case $\Delta =0$.
This allows us to explore rather long chains using a fair number
of disorder realizations (whose magnetization properties on the
other hand, are self-averaging). In Fig (\ref {fig1}) we show
respectively the whole magnetization curves obtained for various
disorder concentrations $p=0,0.2,0.4,0.6,0.8$ and $1$ averaging on
$ 5\times 10^{4}$ samples of $L=f(18)=2584$ sites under the
exchange disorder ( \ref{q-bin}), with $\delta =0.95$ and $U=0.2$.
It can be readily verified that a set of robust plateaux
emerges quite precisely at the critical magnetizations given by
Eq.\ (\ref{dec1A>B}) for the plateau closest to saturation and by
Eq.\ (\ref{dec2A>B}) for the second one.

\begin{figure}[tbp]
\hbox{%
\epsfxsize=3.5in
\epsffile{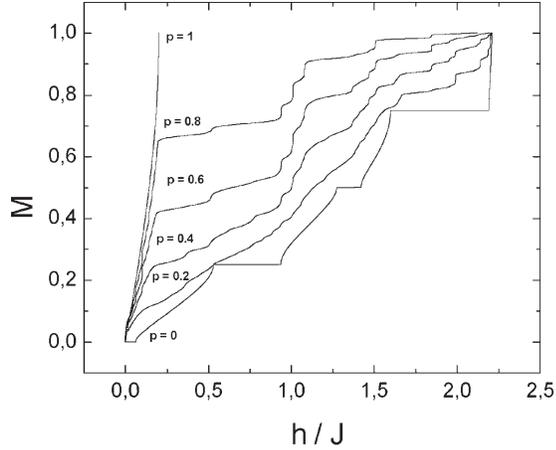}}
\caption{Double frequency magnetization curves of the XX chain for
$\protect \omega _{1}=5/8$ and $\protect\omega _{2}=7/8$ with
amplitudes $\protect \delta _{1}=0.2$ and $\protect\delta
_{2}=0.3$, with $U=0.1$ and $10^{4}$ spins over $100$ samples,
immersed in disordered binary backgrounds of strength
$p=0,0.2,0.4,0.6,0.8$ and $1$ in ascending order} \label{fig2}
\end{figure}

In a previous work \cite{nos1}, we have observed that the
magnetization curve for the Fibonacci chain, could be well
approximated by considering a rather small subset of the main
frequencies in its Fourier spectrum. Here we study a two-frequency
case, for $\omega _{1}=5/8$ and $\omega _{2}=7/8$, in the presence
of disorder, where it is observed that the plateaux are erased
even by small disorder (see Fig.\ (\ref{fig2})). It is interesting
to observe that in contrast to the Fibonacci situation studied
above, in which the plateaux structure is robust and just shifts
with the strength of the disorder, here the plateaux seem to smear
out even for a small value of $p$.

\begin{figure}[tbp]
\hbox{%
\epsfxsize=3.5in
\epsffile{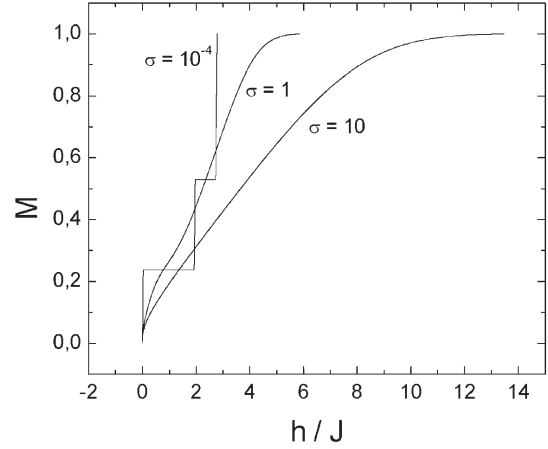}}
\caption{Magnetization curves of modulated XX Fibonacci spin
chains, immersed in Gaussian exchange distributions, after
averaging over $4\times 10^{4}$ samples with $f(18)=2584$ sites,
$\protect\delta =0.95$ and increasing standard deviation from left
to right (note that the leftmost is practically the pure Fibonacci
case).} \label{fig3}
\end{figure}

It is important to stress that the derivation of our results for
the quantization conditions Eqs.\ (\ref{decB>A})-(\ref{dec2A>B})
rely strongly on the discreteness of the probability distribution
and would not to be applicable to an arbitrary continuous exchange
disorder. In fact, for the Gaussian case referred to above it
turns out that no traces of plateaux can be observed. This is
corroborated in Fig.\ (\ref{fig3}), where we see that the plateaux
structure is smoothed when the standard deviation is increased.
Here the sampling was increased up to $4 \times 10^{4}$
realizations though the length of the chain was reduced to
$L=f(18)$, as the CPU time per spectrum grows as $L^{2}$. However,
preliminary computations using larger chains yielded no
substantial differences.

For homogeneously disordered chains, one can use the decimation
procedure of \cite{FDM} along with the universality of the fixed
point, to show that either for discrete or continuous
distributions the low field magnetic susceptibility behaves
according to

\begin{equation}
\chi _{z}\propto \frac{1}{H[\ln (H^{2})]^{3}}\ .  \label{chi-odd}
\end{equation}

\begin{figure}[tbp]
\hbox{%
\epsfxsize=3.5in
\epsffile{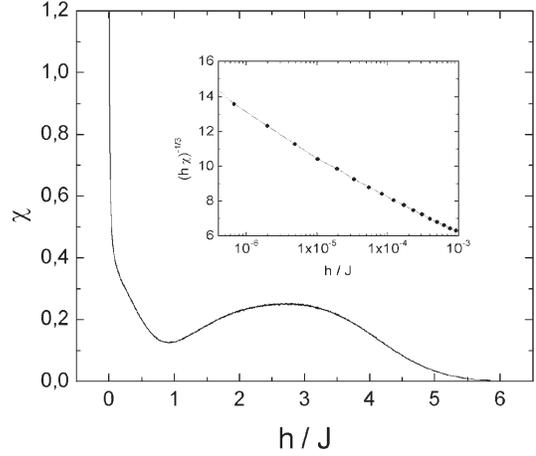}}
\caption{Magnetic susceptibility of modulated XX Fibonacci spin
chains, immersed in a Gaussian exchange distribution
$(\protect\sigma =1)$, after averaging over $4\times 10^{4}$
samples with $f(18)=2584$ sites, $\protect \delta =0.95$. The
inset show the susceptibility behavior at low magnetic fields
which follows closely the logarithmic regime predicted in the
text.} \label{fig4}
\end{figure}

\begin{figure}[tbp]
\hbox{%
\epsfxsize=3.5in
\epsffile{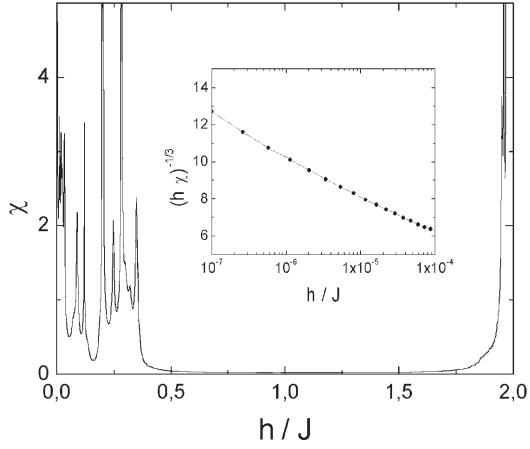}}
\caption{Magnetic susceptibility of modulated XX Fibonacci spin
chains, immersed in a binary exchange disorder of strength $p=0.6$
averaged over $ 5\times 10^{4}$ samples with $f(18)=2584$ sites,
$\protect\delta =0.95$. The inset show the susceptibility behavior
at low magnetic fields that like the Gaussian disorder follows
closely the logarithmic regime.} \label{fig5}
\end{figure}

Following a simple argument based on random walk motion used in
\cite{EggRie} , it can be readily shown that for $\Delta =0$ (or
$XX$ chains), these arguments can be extended to the case of a
disordered Fibonacci chain. It is interesting to note that the
effect of the disorder is crucial since it changes the power law
behaviour of the free Fibonacci chain obtained in \cite {LN} to a
logarithmic one. This can be clearly observed in the insets of
Figs.\ (\ref{fig4}), (\ref{fig5}) where the validity of these
arguments seem to apply over more than two decades.

To summarize, we have studied the effect of disorder on the
plateaux structure in quasiperiodic $XXZ$ chains under an external
magnetic field. By means of a simple real space decimation
procedure we found the values of the magnetization for which
the main plateaux emerge, Eqs.\ (\ref{decB>A})-(\ref{dec2A>B}).
This was tested by numerical diagonalizations of large $XX$ chains
finding a remarkable agreement with the quantization conditions
in a variety of scenarios. Since the decimation scheme applies
for generic $XXZ$ chains \cite{nos2}, we conclude that the appearance
of these plateaux is a generic feature, at least with an antiferromagnetic
anisotropy parameter $0 < \Delta < 1$. This issue still awaits numerical
confirmation on sufficiently long chains using state of the art
methodologies such as density matrix renormalization group \cite{Karen}.
Finally, we have also studied the low magnetic field susceptibility
which exhibits a clear logarithmic behavior, Eq.\ (\ref {chi-odd}).
We trust this work will convey a motivation for both experimental
and numerical studies.

It is a pleasure to acknowledge useful discussions with D.C.\
Cabra and M.D.\ Grynberg. Financial support from Fundaci\'{o}n
Antorchas, Argentina (grant No.\ A-13622/1-106) is acknowledged.

\newpage

\hbox{}

\hbox{}

\hbox{}

\hbox{}

\hbox{}

\end{document}